\documentclass{article}
\usepackage{xcolor}
\usepackage{mdframed}
\usepackage{booktabs}
\usepackage{multirow}
\usepackage{graphicx}
\usepackage{geometry}
\usepackage[colorlinks=true, linkcolor=black, urlcolor=blue, citecolor=blue, anchorcolor=blue]{hyperref}
\usepackage[style=numeric, natbib]{biblatex}
\usepackage{tabularx} 

\graphicspath{{figs/}}

\newcommand{\upstairs}[1]{\textsuperscript{#1}}
\newcommand{\corr}{$\mathsection$}
\newcommand{\chatgptbox}[3]{%
  \begin{mdframed}[
    frametitle={#1},
    frametitlebackgroundcolor=gray!20,
    frametitlerule=true,
    frametitlerulecolor=black,
    frametitlerulewidth=1pt,
    linewidth=1pt,
    linecolor=black,
    frametitlealignment=\centering,
    backgroundcolor=white,
    innertopmargin=10pt,
    innerbottommargin=10pt,
    leftmargin=0pt,
    rightmargin=0pt,
    skipabove=10pt,
    skipbelow=10pt,
    nobreak=true,
  ]
  {\small \noindent\textbf{Prompt:}} {\small #2}
  \vspace{0.2em}
  
  {\small \noindent\textbf{ChatGPT:}} {\small #3}
  \end{mdframed}
  \vspace{1em}
}
\newcommand{\chatgptconverbox}[5]{%
  \begin{mdframed}[
    frametitle={#1},
    frametitlebackgroundcolor=gray!20,
    frametitlerule=true,
    frametitlerulecolor=black,
    frametitlerulewidth=1pt,
    linewidth=1pt,
    linecolor=black,
    frametitlealignment=\centering,
    backgroundcolor=white,
    innertopmargin=10pt,
    innerbottommargin=10pt,
    leftmargin=0pt,
    rightmargin=0pt,
    skipabove=10pt,
    skipbelow=10pt,
    nobreak=true,
  ]
  {\small \noindent\textbf{Prompt:}} {\small #2}
  \vspace{0.2em}
  
  {\small \noindent\textbf{ChatGPT:}} {\small #3}
  
  \vspace{0.2em}

  {\small \noindent\textbf{Prompt:}} {\small #4}
  \vspace{0.2em}
  
  {\small \noindent\textbf{ChatGPT:}} {\small #5}
  \end{mdframed}
  \vspace{1em}
}

\begin{document}

\title{What Should  Data Science Education Do  with Large Language Models?
}
\date{}
\maketitle

\begin{center}
  \vspace{-0.5in}

  \begin{tabular}{cc}
    Xinming Tu \upstairs{1}, James Zou\upstairs{2}, Weijie J.~Su \upstairs{3}\upstairs{\corr}, Linjun Zhang\upstairs{4}\upstairs{\corr}
   \\[0.25ex]
   {\small \upstairs{1} University of Washington} \\
   {\small \upstairs{2} Stanford University} \\
   {\small \upstairs{3} University of Pennsylvania} \\
   {\small \upstairs{4} Rutgers University} \\
      {\small \upstairs{\corr} Corresponding authors} \\
  \end{tabular}

\end{center}

\vspace*{0.15in}
\hspace{10pt}
  

\fontsize{11pt}{13.2pt}\selectfont

\begin{abstract}
The rapid advances of large language models (LLMs), such as ChatGPT, are revolutionizing data science and statistics. These state-of-the-art tools can streamline complex processes such as data cleaning, model building, interpretation, and report writing. As a result, it reshapes the role of data scientists.
We argue that LLMs are transforming the responsibilities of data scientists, shifting their focus from hands-on coding, data-wrangling and conducting standard analyses to assessing and managing analyses performed by these automated AIs. This evolution of roles is reminiscent of the transition from a software engineer to a product manager, where strategic planning, coordinating resources, and overseeing the overall product life cycle supersede the task of writing code. We illustrate this transition with concrete data science case studies using LLMs in this paper.

These developments necessitate a meaningful evolution in data science education. Pedagogy must now place greater emphasis on cultivating diverse skillsets among students, such as LLM-informed creativity, critical thinking, AI-guided programming, and interdisciplinary knowledge. LLMs can also play a significant role in the classroom as interactive teaching and learning tools, contributing to personalized education and enriched learning experiences. This paper discusses the opportunities, resources and open challenges for each of these directions. As with any transformative technology, integrating LLMs into education calls for careful consideration. While LLMs can perform repetitive tasks efficiently,  it's crucial to remember that their role is to supplement human intelligence and creativity, not to replace it. Therefore, the new era of data science education should balance the benefits of LLMs while fostering complementary human expertise and innovations. In conclusion, the rise of LLMs heralds a transformative period for data science and its education. This paper seeks to shed light on the emerging trends, potential opportunities, and challenges accompanying this paradigm shift, hoping to spark further discourse and investigation into this exciting, uncharted territory.
\end{abstract}

\newpage


\section{Introduction}

The rapid advancements in artificial intelligence have led to the development of powerful tools, one of the most notable being Large Language Models (LLMs) such as ChatGPT by OpenAI \citep{brown2020language, OpenAI_2023}. These models have demonstrated remarkable capabilities in understanding and generating human-like text, often outperforming traditional algorithms in various natural language processing tasks. The rise of LLMs has brought forth a paradigm shift in the field of data science and has the potential to reshape the way we approach data science education. This article will focus on the impact of LLMs in this field. 

The role of Data Scientists has been heralded as the "sexiest job of the 21st century" by the Harvard Business Review \citep{davenport2012, davenport2022}. This is due to the explosive growth of digital information, leading to the necessity for expertise in data-driven domains such as health care, advertisement recommendation, and job applications \citep{mayer2013big}. Data science education aims to equip students with the knowledge and skills required in these rapidly evolving fields. The advent of LLMs further revolutionizes this landscape, demanding a shift in both the content of data science education (\textit{what to teach/learn}) and the methods of data science education (\textit{how to teach/learn}). It is incumbent upon educators and students alike to recognize and adapt to the transformative power of LLMs in this new era.

The emergence of LLMs such as OpenAI's GPT-4 marks a transformative shift across numerous industries, most notably data science \citep{Eloundou_2023}. Recent findings further demonstrate GPT-4's impressive capabilities, showcasing performance on par with humans across a variety of data analysis tasks \citep{cheng2023gpt4}. By automating complex processes, streamlining code generation, and facilitating role transitions, LLMs possess the potential to redefine not only the data science pipeline but also the fundamental nature of data science education. In this new era of LLMs,  students need to learn to view themselves as \textit{product managers} rather than \textit{software engineers}, that is, their focus should be shifted to strategic planning, coordinating resources, and overseeing the overall product life cycle, rather than the standard data analysis pipeline.

This paper will provide a holistic examination of the transformative potential of LLMs on the data science pipeline, utilizing a heart disease dataset to illustrate the capabilities of the ChatGPT-plugin, an LLM equipped with a code plugin. The model performs tasks ranging from data cleaning and exploration to model building, interpretation, and report writing, thus demonstrating its impressive adaptability and problem-solving capabilities. The role of LLMs in enhancing various stages of the data science pipeline and redefining the responsibilities of data scientists will be explored, along with the shifting emphasis in data science education towards a diverse skillset encompassing creativity, critical thinking, LLM-guided programming, and interdisciplinary knowledge.

Following this, we will examine the integration of LLMs into data science education. From curriculum design to personalized tutoring and the development of automated education systems, LLMs offer numerous possibilities to enrich the teaching and learning experience. Educators can leverage LLMs to design dynamic curricula, generate contextually relevant examples, and stay abreast of industry trends. Furthermore, as powerful teaching assistants, LLMs can provide personalized guidance to students, leading the way to a more engaging and interactive learning environment.

The structure of this paper is as follows: we start with an overview of the current state of LLMs and data science education, followed by a discussion on the impact of LLMs on data science and the need to redefine its content to prepare students for the paradigm shift. We then explore the potential of LLMs as interactive teaching and learning tools, envisioning an automated education system that fosters personalized learning experiences. Subsequently, we delve into the necessary precautions and considerations when integrating LLMs into the educational system, highlighting the balance of utilizing LLMs to reduce repetitive tasks while nurturing human intelligence and creativity. Finally, we explore the future of data science education, discussing the potential opportunities and challenges that lie ahead.

\section{Current state of LLMs and Data Science education}

\subsection{Current State of LLMs} 
LLMs represent a powerful class of artificial intelligence models, specifically devised to understand, interpret, and generate human language with exceptional precision. Generative Pretrained Transformers (GPT) stand as one of the most potent LLMs. The fundamental principle underpinning GPT is next-word prediction \citep{radford2018improving}, a seemingly simple concept that catalyzes its extraordinary performance.

The remarkable proficiencies of LLMs can be ascribed to their capacity to process, reason, and learn from vast datasets. These datasets often comprise billions of words and phrases culled from an assorted array of sources, including code repositories, online dialogues, articles, and various other internet resources. This comprehensive training enables LLMs to cultivate an extensive understanding of language~\citep{devlin2019bert}, common sense \citep{moghaddam2023boosting, dhingra2023mind}, reasoning \citep{liu2023evaluating}, showcasing a semblance of intelligence \citep{Bubeck_Chandrase2023}.

OpenAI's recent breakthrough, ChatGPT (based on GPT4), underscores the impressive potential of LLMs in executing a myriad of tasks \citep{Bubeck_Chandrase2023}. This innovation is poised to instigate revolutionary changes across diverse societal facets, including education, programming \citep{Welsh2023}, and the broader labor market \citep{Eloundou_2023}, underscoring the transformative influence of LLMs in steering the future trajectory of artificial intelligence and its practical applications. Furthermore, recent advancements have equipped LLMs with the ability to adapt and utilize various tools, signaling an unprecedented level of capability\citep{shen2023hugginggpt}. For instance, their integration with code interpreters enables LLMs to perform complex coding tasks, including automatic debugging during code generation. Additionally, browsing capabilities equip LLMs with the ability to access up-to-date information, thus enhancing their relevance and practical utility \citep{nakano2022webgpt}.

\subsection{Current State of Data Science Education}

The traditional data science curriculum encompasses a diverse range of subjects aimed at providing students with a strong foundation in the field. Core topics often include statistics, probability, linear algebra, programming (usually with languages Python or R), machine learning algorithms, data visualization, and databases \citep{de2017curriculum,cao2017data}. The curriculum is designed to equip students with the necessary technical skills to collect, analyze, and interpret data, as well as to create and deploy models for various applications such as finance, healthcare, and social sciences.

Teaching methods in data science education typically involve a combination of lectures, labs, and assignments\citep{hicks2018guide}. Lectures provide the theoretical background, introducing students to key concepts and principles. Labs offer practical experience in applying these concepts, often through coding exercises and the use of popular data science libraries and tools. Assignments and projects further reinforce the learning process by challenging students to apply their knowledge to real-world problems, usually involving real or simulated datasets.

\section{The Impact on Data Science Education Content}
As LLMs revolutionize the data science pipeline, their transformative potential is driving significant changes in data science education. This section will concentrate on how these developments are altering the content - or the ``WHAT'' - of data science education. The subsequent section will delve into the evolving methodologies for integrating LLMs into the education system - essentially the ``HOW''. We will begin by examining how LLMs are reshaping the education field, from streamlining various stages of the pipeline to 
solving exam problems. 

\subsection{Transforming the Data Science Pipeline with Large Language Models}\label{sec:pipeline}
LLMs have the potential to revolutionize the data science pipeline by simplifying complex processes, automating code generation, and redefining the roles of data scientists. With the assistance of LLMs, data scientists can shift their focus towards higher-level tasks, such as designing questions and managing projects, effectively transitioning into roles similar to product managers.
\begin{figure}
    \centering
    \includegraphics[width=0.8\textwidth]{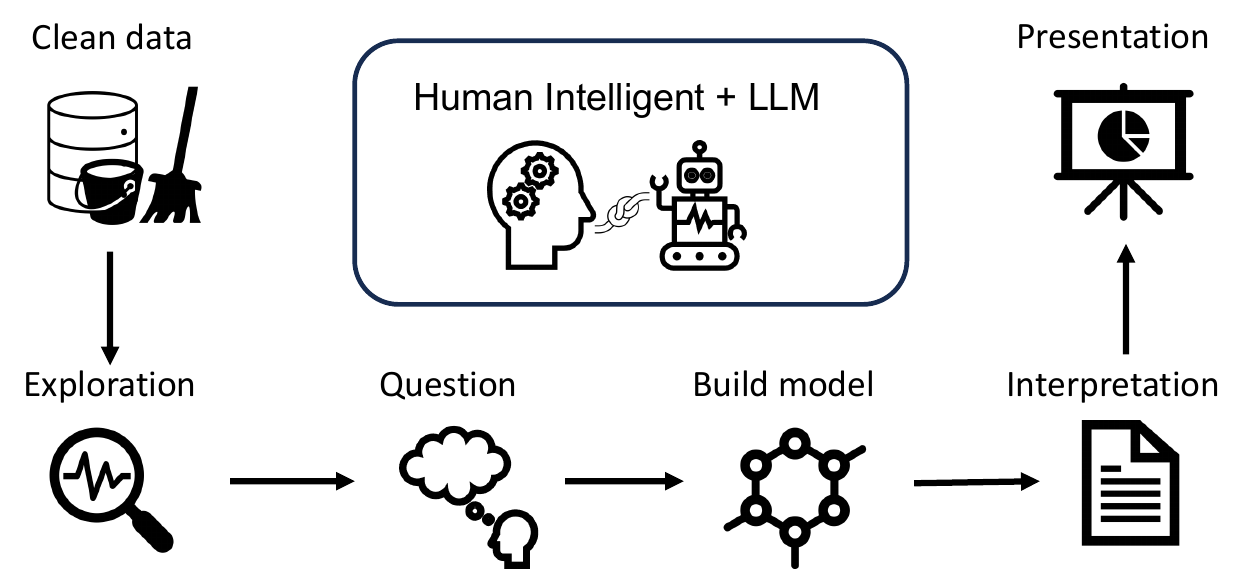}
    \caption{LLMs can revolutionize the entire data science pipeline, from data cleaning and exploration to model building and final presentation. The future pipeline of the future data science is the collaboration between Human intelligent and LLMs.
    }
    \label{fig:LLM_DS_scheme}
    \vspace{-10pt}
\end{figure}
In our following case study, we will show that LLMs can significantly streamline various stages of the data science pipeline, including:
\begin{itemize}
    \item Data cleaning: LLMs can automatically generate code for cleaning, preprocessing, and transforming raw data, saving data scientists considerable time and effort.
    \item Data exploration: LLMs can generate code for exploratory data analysis, identifying patterns, correlations, and outliers in the data.
    \item Model building: LLMs can suggest appropriate machine learning models based on the problem at hand and generate the necessary code to train and evaluate these models.
    \item Model interpretation: LLMs can help data scientists understand the intricacies of the models they have built, highlighting important features and explaining model behavior in human-readable terms.
    \item Presentation of results: LLMs can generate visuals, reports, and summaries to effectively communicate the findings of a data science project to both technical and non-technical stakeholders.

\end{itemize}

To illustrate the transformative potential of LLMs in the data science pipeline, let us consider the following example:

We use a \href{https://www.kaggle.com/fedesoriano/heart-failure-prediction}{heart disease dataset} on Kaggle \citep{kaggle2021}, which contains records of individuals with various cardiovascular risk factors and diagnostic information. The primary objective of this dataset is to scrutinize the correlation between these risk factors and heart disease, as well as to construct a predictive model for heart disease. This dataset was posted on Kaggle after September 2021. The training data for ChatGPT (gpt-4, gpt-3.5-turbo) only extends up to September 2021 based on the system prompt.

Our goal is to perform a data-science pipeline analysis of this dataset using the ChatGPT code plugin, which can interact with a Python interpreter so that it can run the generated code. By providing just a few prompts, we aim to accomplish tasks such as data cleaning, data exploration, model building, model interpretation, and ultimately, report writing.

We summarize our results in Table \ref{tab:chatgpt_actions}. It is essential to note that the left column, labeled ``Prompt", contains all text inputs (excluding \textit{continue} and \textit{do that} confirmation responses). The right column, labeled ``ChatGPT with Code Interpreter" lists all the tasks completed by ChatGPT using the code plugin. As illustrated by the results, with only simple prompts (comprising a few words), ChatGPT is capable of completing the entire data analysis pipeline. The detailed prompts and the complete conversation history with ChatGPT are deferred to the appendix. 

\begin{table}
\centering


\begin{tabular}{@{}ll@{}}
\toprule
\textbf{Prompt input} & \textbf{Finished Task by ChatGPT with Code Interpreter} \\ \midrule
\multirow{4}{*}{Do some data cleaning} & Check for Missing/Null values \\
 & Remove duplicate rows if necessary \\
 & Check for inconsistent/invalid values \\
 & Convert categorical columns to numerical representations \\ \midrule
\multirow{4}{*}{Do some data explorations} & Compute summary statistics \\
 & Create distribution plots \\
 & Compute correlation matrix \\
 & Analyze heart disease prevalence \\ \midrule
\multirow{4}{*}{Build a model to predict Heart Disease} & Split data into training/testing sets \\
 & Train logistic regression model \\
 & Evaluate model performance \\
 & Interpret results \\ \midrule
\multirow{2}{*}{Use better models} & Try Random Forest, Support Vector Machine, \\
 & Gradient Boosting \\
 & Report accuracies for each model \\ \midrule
Plot performance for these models & Create bar plots and comparison of model performances \\ \midrule
\multirow{4}{*}{Use the best model and try to improve it} & Perform grid search \\
 & Evaluate with cross-validation \\
 & Reduce search space \\
 & Get the final model \\ \midrule
Plot performance for different parameters & Create plots of performance for different parameters \\ \midrule
Interpret the best model & Obtain feature importance and explain the results \\ \midrule
Write a report for this project & Write a comprehensive report \\ \bottomrule
\end{tabular}
\caption{Summary of Tasks Completed by ChatGPT Based on Prompts. The first column represents the input prompts provided to ChatGPT, while the second column summarizes the tasks accomplished by ChatGPT in response to these prompts. It is important to note that, for demonstration purposes, we have minimized the input to showcase the remarkable capabilities of ChatGPT}

\label{tab:chatgpt_actions}
\end{table}

We highlight a few selective tasks to demonstrate both the figures and code generated by ChatGPT, which include data exploration, model building, hyperparameter search, model interpretation, and report writing.

\begin{figure}
    \centering
    \includegraphics[width=0.9\textwidth]{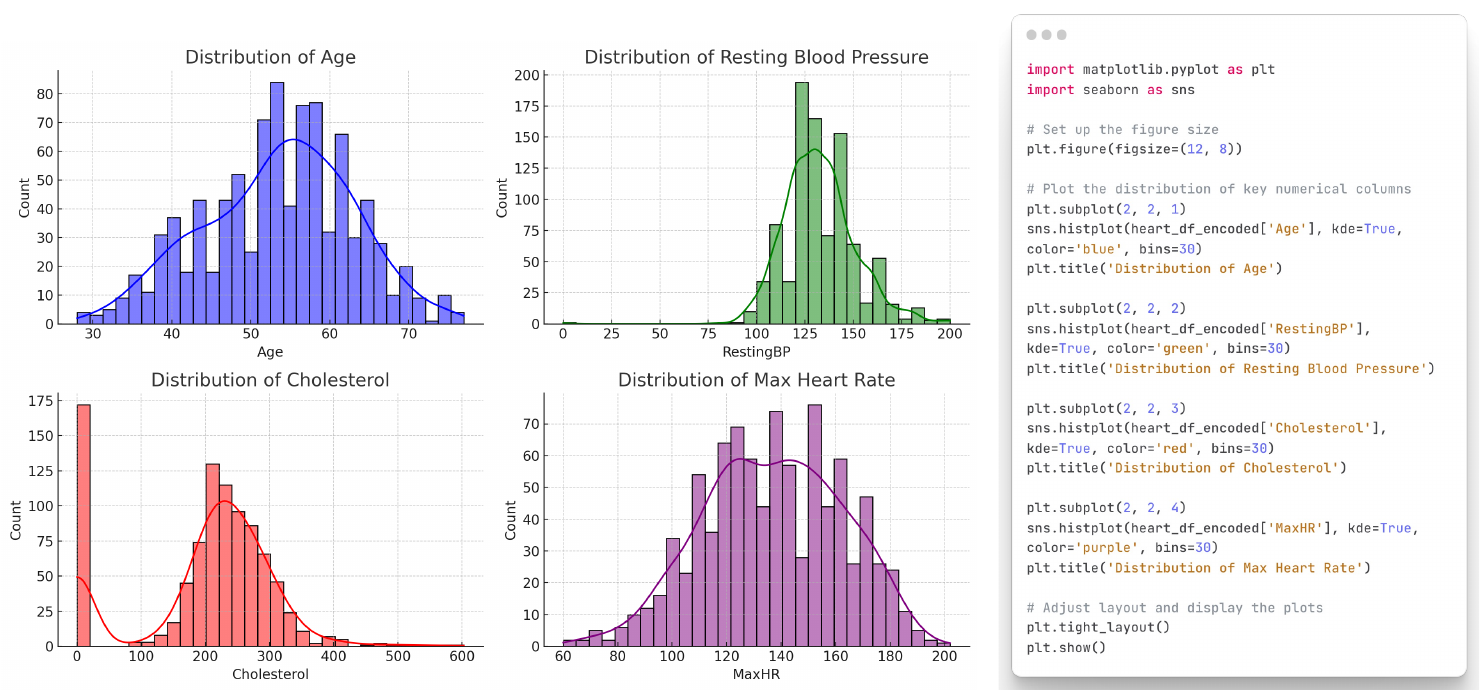}
    \caption{ChatGPT's capabilities in generating code for data exploration, represented by the derived distribution plots. On the left are four distribution plots that ChatGPT generated, while on the right is a snapshot of the code used to generate these plots.
    }
    \label{fig:explore_dist}
    \vspace{-10pt}
\end{figure}

\begin{figure}
    \centering
    \includegraphics[width=0.9\textwidth]{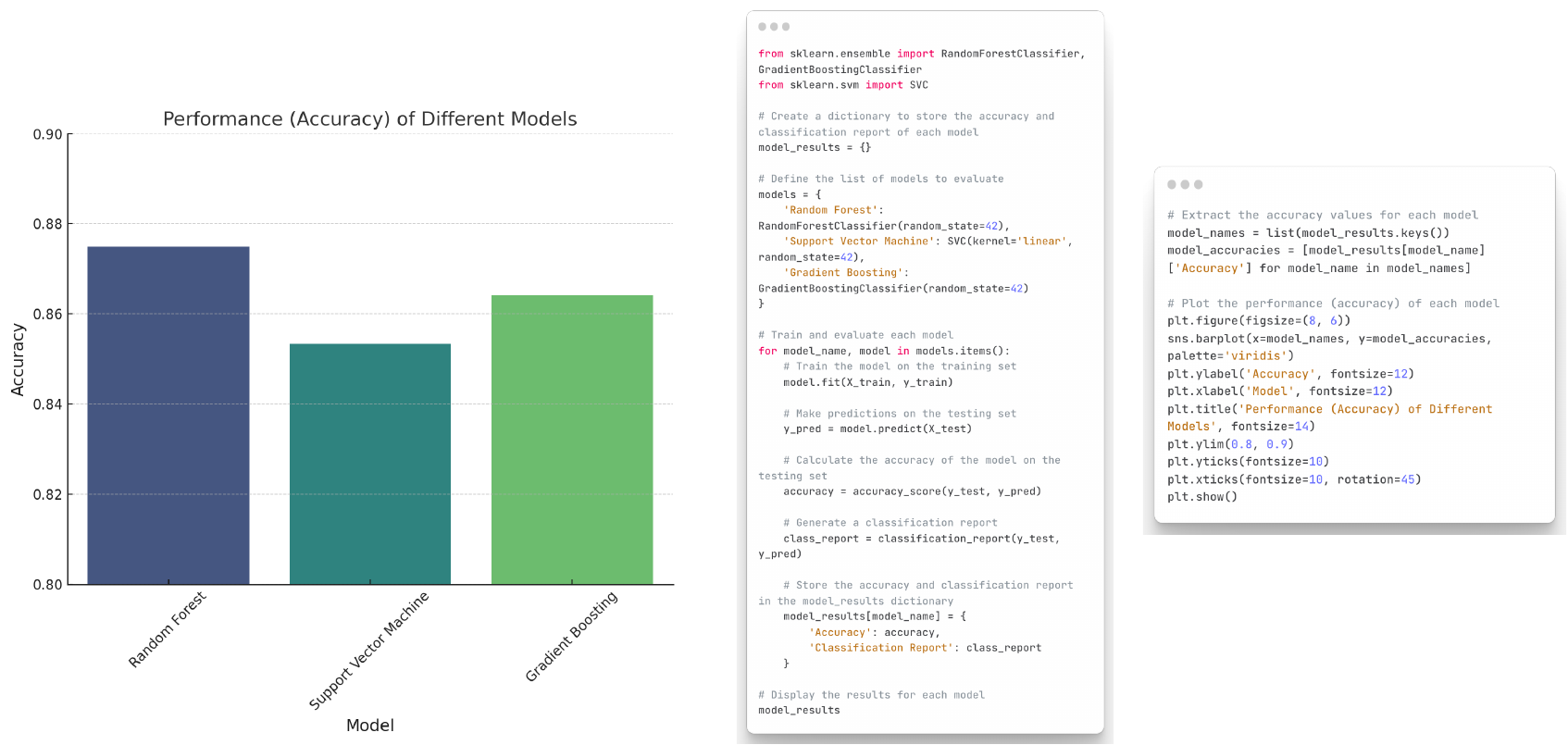}
    \caption{Building multiple models: ChatGPT tried Random Forest, Support Vector Machine and Gradient Boosting method and compare the performances. On the left are bar plots depicting the predictive performance of different models. On the right is a snapshot of the code used to generate these plots.
    }
    \label{fig:build_models}
    \vspace{-10pt}
\end{figure}

For instance, when given the prompt \textit{do some data explorations}, ChatGPT produces distribution plots for data exploration (Figure \ref{fig:explore_dist}). 
Upon the prompt \textit{Use better models}, ChatGPT employs Random Forest, Support Vector Machine, and Gradient Boosting, and then plot the bar chart to compare their prediction performance (Figure \ref{fig:build_models}).

\begin{figure}
    \centering
    \includegraphics[width=0.9\textwidth]{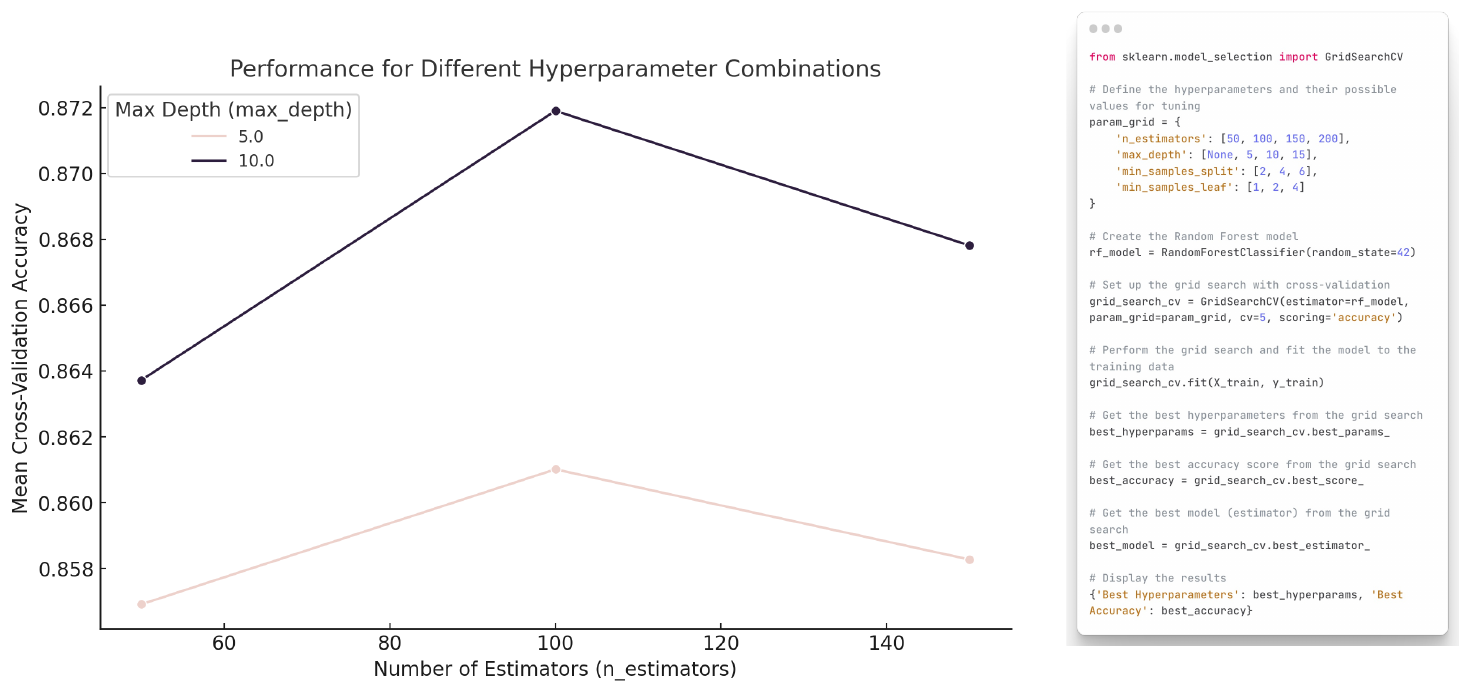}
    \caption{Hyperparameter Optimization: ChatGPT also demonstrates the ability to search the hyperparameter space to identify optimal models. On the left are the mean cross-validation accuracy corresponding to different hyperparameters. On the right, a code snapshot reveals the process employed to generate these plots.
    }
    \label{fig:hyper_search}
    \vspace{-10pt}
\end{figure}

\begin{figure}
    \centering
    \includegraphics[width=0.9\textwidth]{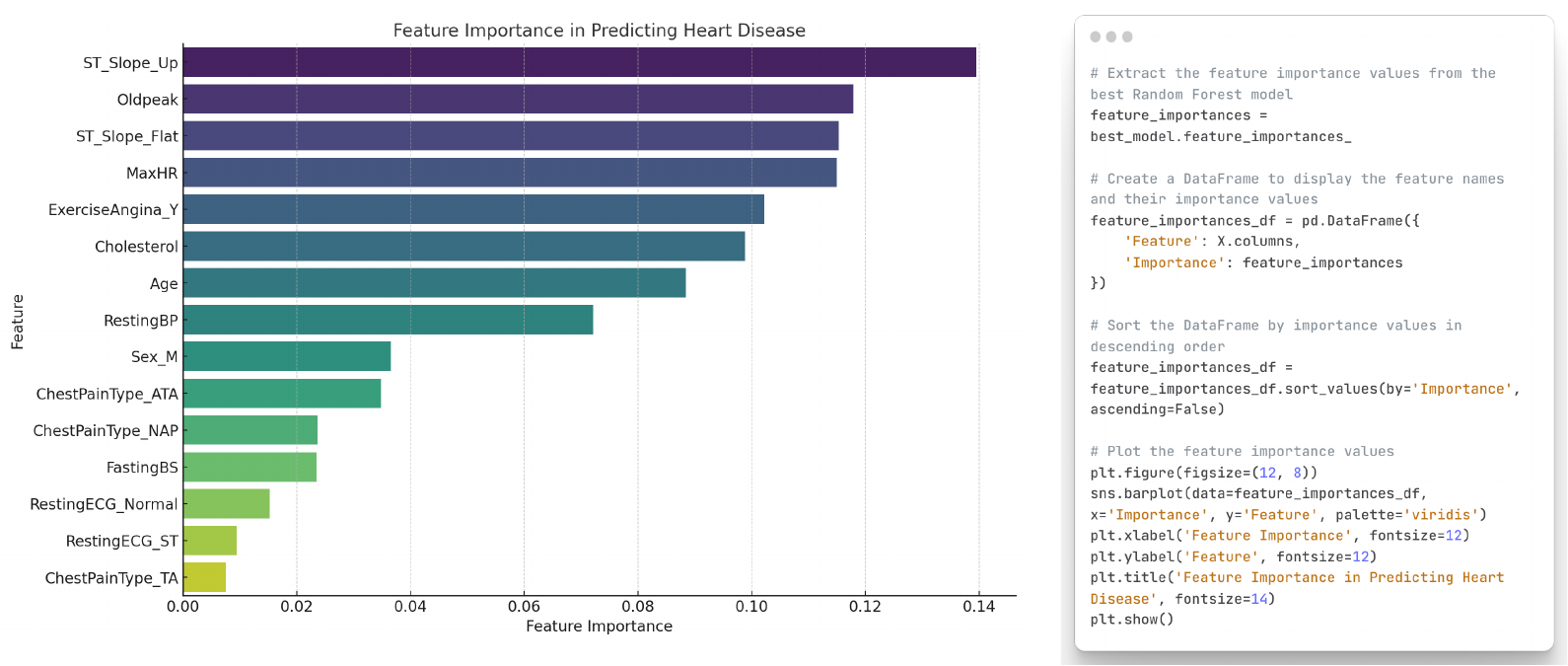}
    \caption{Explaining the models: With the prompt "interpret the best model", ChatGPT can use the feature importance score to explain the model and give the plots to show which feature is more important. Left: feature importance barplots. Right: Code snapshot to generate the plots.
    }
    \label{fig:interpret_model}
    \vspace{-10pt}
\end{figure}

Furthermore, ChatGPT can improve the best performing model by executing a hyperparameter search. In response to this task, it autonomously defines the search space and identifies the best model (Figure \ref{fig:hyper_search}). With the prompt \textit{interpret the best model}, ChatGPT utilizes feature importance scores to explain the model, generating plots to illustrate the significance of each feature (Figure \ref{fig:interpret_model}). Finally, using the prompt \textit{write a report for this project}, ChatGPT produced a draft for a project report that encapsulates all the previous sections. Though the output context has its limitations and lacks granular details, it nonetheless provides a satisfactory report of the project.

Impressively, when ChatGPT encounters errors, it can auto-debug based on the error output information and revise the code by itself. Furthermore, when conducting hyperparameter search, if ChatGPT finds that the process takes longer than expected (resulting in a timeout error), it can intelligently learn to reduce the search space. For a detailed view with figures of the \href{https://chat.openai.com/share/daecaa84-a8d7-47cb-bc2c-93b0c36eaa1a}{conversation history}, please refer to the supplementary material. This level of adaptability demonstrates the remarkable capabilities of ChatGPT in implementing the data science pipeline.  

\subsection{Exam-Taking Abilities of ChatGPT}\label{sec:exam}
In this subsection, we conduct an evaluation of ChatGPT on statistical examinations, which include both conceptual and coding problems. For this purpose, we sourced exercises from ``Introduction to Statistical Thinking'' \citep{yakir2011introduction}, spanning fifteen chapters. This book, being not as widely used, minimizes the risk of data leakage, and its original solutions are provided in R. In contrast, ChatGPT produces solutions in Python, serving to underscore its generalized performance. 

\begin{figure}
    \centering
    \includegraphics[width=0.8\textwidth]{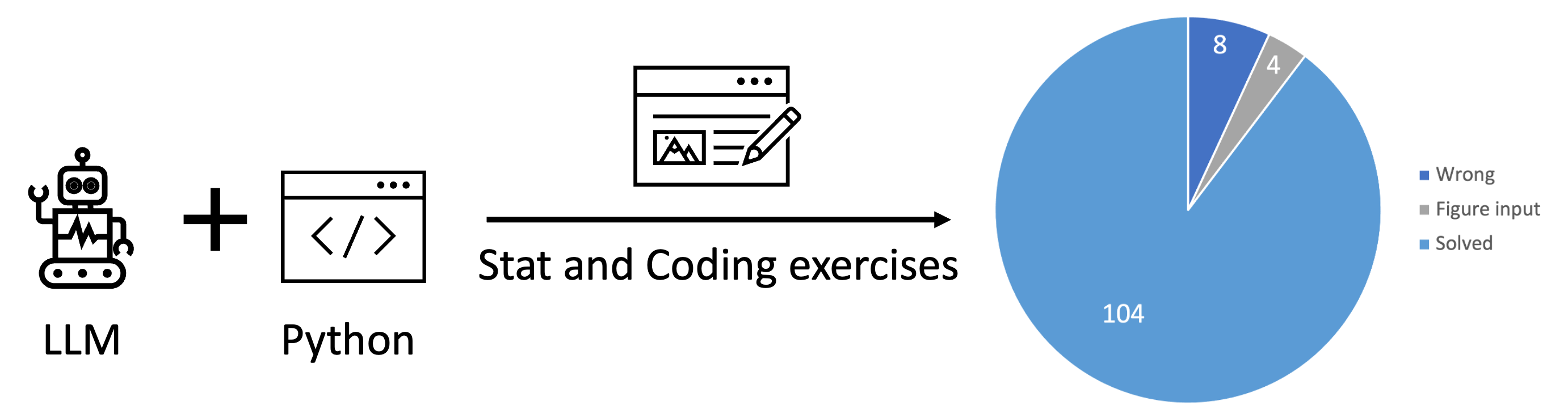}
    \caption{With the code plugin, ChatGPT was able to solve 104 out of 116 exercises in the book "Introduction to Static Thinking." However, in the 4 exercises that required a figure as input, ChatGPT encountered difficulties and made mistakes in the remaining 8 exercises.
    }
    \label{chat_stat}
\end{figure}

We converted all problems (especially equations) into a LaTeX version, and the problem statement in LaTeX was used as the input without any prompt engineering. The solutions provided by ChatGPT were then compared with the book's solutions, and points were manually assigned based on this comparison. Our results revealed that ChatGPT exhibited an impressive performance, securing 104 points out of a total of 116. The model could not answer some questions (totaling 4 points) that required the interpretation of a figure as input (Figure \ref{chat_stat}). The main failures were predominantly due to the model misunderstanding the questions, such as don't know how to interpret the input format of the table. Detailed conversations related to these issues can be found in the supplemental material. Recent research has suggested that ChatGPT exhibits promising results, often resembling human-like performance in various data analysis tasks \citep{cheng2023gpt4}. It's essential to bear in mind that our results are not indicative of the model's superiority or equivalence to human proficiency. They serve to demonstrate that students could potentially misuse advanced AI models like ChatGPT to manipulate standard exam questions. Thus, we must remain vigilant and adapt our educational and evaluation practices accordingly. We will discuss how the education methods should adapt to LLMs in Section~\ref{sec:change}.

As we see from the above two showcases of the remarkable capabilities of LLMs, the traditional data science workflow is on the verge of a paradigm shift. 
LLMs are transforming the roles of data scientists and even the nature of programming itself \citep{Welsh2023}. With the increased integration of LLMs into the data science pipeline, future data scientists will be able to delegate routine tasks to LLMs, allowing them to focus more on higher-level tasks such as designing questions and managing projects.  Having illustrated what LLMs can do in data science education content, in the next section, we will discuss how to integrate LLMs into data science education to enhance teaching efficacy.

\section{Using Large Language Models in Data Science Education}

While LLMs have a significant impact on data science education, they also present new opportunities for integration and enhancement of teaching efficacy. As state-of-the-art natural language processing tools, LLMs offer a wealth of possibilities for enriching the teaching and learning experience, revolutionizing the way educators and students approach data science. This section explores the incorporation of LLMs into data science teaching, focusing on two key aspects: curriculum design and personalized tutoring. 

\subsection{Design curriculum with LLMs}
First, we discuss how LLMs can assist educators in designing dynamic and engaging curricula, generating contextually relevant examples, exercises, and explanations that help students grasp complex concepts with greater ease. 
Importantly, LLMs enable educators to stay up-to-date with the latest advancements and industry trends, ensuring that their courses remain relevant and aligned with the evolving needs of the data science field.
\chatgptconverbox{\href{https://chat.openai.com/share/4b47fda4-337d-4bd4-8b22-2d2a4925bf21}{Example of customized curriculum design}}{I want to teach my students about the data science, they are not familiar with the python, help me design the curriculum}{Absolutely, I'd be happy to help you design a curriculum for teaching data science to students unfamiliar with Python. This curriculum will first introduce students to the basics of Python programming, and then delve into core data science concepts and skills.
\begin{itemize}
    \item Introduction to Data Science
    \item Introduction to Programming with Python
    \item Data Manipulation and Analysis
    \item Data Visualization
    \item Statistical Analysis and Hypothesis Testing
    \item Machine Learning Fundamentals
    \item Machine Learning Models and Evaluation
    \item Introduction to Deep Learning
    \item Data Science Project
\end{itemize}
Please note that this is a very generalized curriculum. Depending on the background and pace of your students, you might need to adjust the contents and speed. The idea is to start from basics and gradually move to more complex topics.}{All students are studying biology, could you please revise the curriculum}{Certainly! Considering that the students are studying biology, here's a revised curriculum that focuses on applying data science techniques to biological data:
\begin{itemize}
    \item Introduction to Data Science in Biology
    \item Python Basics for Biology
    \item Exploratory Data Analysis in Biology
    \item Machine Learning in Biology
    \item Deep Learning in Biology
    \item Genomics and Bioinformatics
    \item Data Science Project in Biology
\end{itemize}
}
In the example above, we engage ChatGPT to develop a curriculum for data science education catered to students without a Python background. ChatGPT, in response, constructs a comprehensive teaching plan for data science. Further, when we specify that the students' primary field of study is biology, ChatGPT modifies the curriculum to suit their needs. Due to length limitations, only the module titles from the conversation are included here; the complete conversation, encompassing detailed content for each module, can be found in the supplementary material. 

Furthermore, ChatGPT can be used to aid in quiz design. In the following example, we utilize ChatGPT to generate ten specific questions concerning hypothesis testing.

\chatgptbox{\href{https://chat.openai.com/share/ce8540fd-5036-4112-9a49-8d3178485998}{Example of quiz generation}}{give me 10 specific questions about the hypothesis testing}{
\begin{enumerate}
\item What is the null hypothesis that you are trying to test in your study?
\item What is the alternative hypothesis in your research and how does it relate to the null hypothesis?
\item What statistical test (e.g., t-test, chi-square test, ANOVA) are you using for your hypothesis testing and why did you choose this test?
\item What significance level ($\alpha$) have you chosen for your hypothesis testing and why?
\item Have you performed any assumption checks (such as checking for normality or homogeneity of variance) before conducting your hypothesis test? If so, what were the results?
\item If there are any violations of the assumptions of your chosen statistical test, how are you addressing these in your analysis?
\item How are you addressing the possibility of Type I or Type II errors in your study?
\item What is the power of your test, and is it adequate to detect the effects you are interested in?
\item What are the results of your hypothesis test (p-value), and how do you interpret these results in the context of your study?
\item Based on your hypothesis testing results, what conclusions are you drawing about your research question, and what are the implications for future research?
\end{enumerate}}

\subsection{LLMs as Teaching Assistants}
Large language models (LLMs) have the potential to significantly impact education by serving as powerful teaching tools and customized tutors, assisting students throughout their learning journey. The 2 Sigma Problem, a phenomenon observed by educational researcher Benjamin Bloom, highlights that personal tutors can lead to a two-standard deviation improvement in student performance compared to traditional group instruction \citep{bloom19842}.

We have compiled a list of current GPT-powered tools for coding, which is summarized in Table~\ref{table:AI_Tools_Functions}. These tools offer capabilities such as assisting students with writing code, debugging, and interpretation. These AI-enabled resources serve as teaching assistants, providing hints or even direct guidance on coding tasks. This approach significantly enhances the programming learning experience for students.

Furthermore, in the foreseeable future, we anticipate that LLMs will serve as virtual teaching assistants in broader contexts. These AI-powered tutors can respond to student questions, clarify complex concepts, and provide tailored recommendations for further study. By fostering an engaging and interactive learning environment, LLMs can help students overcome challenges and accomplish their educational objectives. The use of LLMs in education has the potential to narrow the performance gap identified by Bloom \citep{bloom19842}, making personalized learning experiences more accessible and efficient. As a toy example, in the following illustration, when the student wants to know more about A/B test, ChatGPT nicely explains the concept, and offers an example to help understanding.

\begin{table}[ht]
\centering
\begin{tabularx}{\textwidth}{>{\hsize=0.2\hsize}X>{\hsize=0.8\hsize}X}
\toprule
\textbf{Tool Name} & \textbf{Function} \\
\midrule
\href{https://github.com/features/copilot}{Github Copilot} & Uses OpenAI Codex to suggest real-time code and complete functions directly in the editor. \\
\href{https://github.com/features/preview/copilot-x}{Github Copilot X} & AI-powered software development tool featuring chat and terminal interfaces, pull request support, and integration with OpenAI’s GPT-4. \\
\href{https://www.cursor.so}{Cursor} & A ChatGPT-styled AI programming integrated development environment (IDE). \\
\href{https://openai.com/blog/chatgpt-plugins}{ChatGPT Plugin Code Interpreter} & A version of the ChatGPT model that can interpret Python and manage file uploads and downloads. \\
\href{https://openai.com/blog/chatgpt-plugins}{ChatGPT Plugin Wolfram} & A variant of ChatGPT that leverages Wolfram Language for programming and problem-solving. \\\href{https://python.langchain.com/docs/get_started/introduction.html}{LangChain} & Provides a standardized memory interface accompanied by a series of memory system implementations. It facilitates the execution of more complex tasks utilizing the memory system. \\
\bottomrule
\end{tabularx}
\caption{Overview of Various AI Tools and Their Functions.}
\label{table:AI_Tools_Functions}
\end{table}

\chatgptbox{\href{https://chat.openai.com/share/95bb4504-19d2-4550-a36b-3632eb76a15e}{Example of personal tutor}}{what's the A/B test? Do you have any examples?}{
An A/B test, also known as split testing or bucket testing, is a method of comparing two or more versions of a web page, advertisement, or other digital content to determine which one performs better. The goal is to improve user engagement, conversion rates, or other key performance indicators (KPIs) by identifying and implementing the most effective elements.

During an A/B test, the audience is randomly divided into groups, each of which is exposed to a different version of the content (Version A and Version B). The performance of each version is then measured and compared to determine which one yields better results.

Example:
Imagine you run an e-commerce website and want to increase the number of users who sign up for your newsletter. You hypothesize that changing the design of the newsletter subscription button may impact the sign-up rate.

Version A: This is the control version and uses the current design of the newsletter subscription button, which is a plain text link saying "Subscribe to our newsletter."

Version B: This is the treatment version and uses a new design, featuring a more prominent button with bright colors and the text "Get Exclusive Deals - Subscribe Now!"

You would then randomly assign your website visitors to see either Version A or Version B of the subscription button. After collecting enough data, you would analyze the results to see which version led to a higher sign-up rate. If Version B outperforms Version A, you might decide to implement the new design to boost newsletter sign-ups.}

\section{Adapting Teaching Methods in the New Area}\label{sec:change} 

As we see in the above sections, LLMs exhibit huge potential in benefiting and reshaping data science education. 
In this section, we will discuss the shifting focus in data science education to adapt to the change, emphasizing the need to cultivate a diverse skill set that encompasses creativity, critical thinking, LLM-guided programming, and interdisciplinary knowledge.

\subsection{Teach LLM-Guided Programming}
As highlighted in Section~\ref{sec:pipeline}, LLMs possess the capability to generate and even debug code. Instructing students on leveraging LLMs for programming assistance can notably enhance efficiency.

The process of acquainting students with LLM-guided programming involves incorporating corresponding exercises into the curriculum. This enables students to gain hands-on experience with generating code, resolving debugging issues, and optimizing solutions using LLMs. 

Offering workshops or tutorials on effectively communicating with LLMs to generate desired outputs and refine generated code can enhance students' understanding of LLM-guided programming. 
It's also beneficial to encourage students to explore the capabilities and limitations of LLMs through practical projects. Such an approach showcases the potential of LLMs to expedite the development process and identifies areas where human input remains indispensable. The inclusion of case studies and examples that highlight real-world applications of LLM-guided programming across various industries underlines the burgeoning relevance of this competency.

\subsection{Prevent Plagiarism}
As we show in Sections~\ref{sec:pipeline} and \ref{sec:exam}, LLMs demonstrate exceptional proficiency in handling homework and exams. However, this brings about the need for effective strategies to deter cheating and plagiarism facilitated by LLMs.

A multifaceted approach is required to address this issue. First, assignments should be thoughtfully designed to necessitate critical thinking, personalized reflections, or unique problem-solving approaches that cannot be easily replicated by AI models. For instance, students can be prompted to summarize their learnings from a class or articulate answers in person, further enabling authentic engagement with the course material.
Furthermore, implementing plagiarism detection tools is another crucial strategy. These tools are increasingly sophisticated, often capable of identifying AI-generated content. Nonetheless, recent studies indicate that these detectors sometimes yield false accusations against students \citep{jimenez2023professors} and may display biases towards specific groups \citep{liang2023gpt}. Designing AI detectors that yield low false-positive rates remains a formidable challenge, signifying an area for further exploration.
Finally, it is essential that educators explicitly address academic integrity in the context of these AI tools. By doing so, they can cultivate a clear understanding among students of what constitutes plagiarism and the severe consequences of such unethical practices. Ensuring students' awareness of these issues is a vital step towards promoting responsible and ethical use of LLMs in education.

\subsection{Nurturing Creativity and Critical Thinking}

Although LLMs, such as GPT-4, have shown remarkable capabilities, achieving results comparable to humans on coding platforms like LeetCode and performing well in technical job interviews \citep{Bubeck_Chandrase2023}, it is crucial to recognize their limitations when integrating them into data science education.

One core limitation of GPT-4 arises from its autoregressive nature, which affects the model's ability to plan and think. Autoregressive models like GPT-4 generate outputs one token at a time, conditioning each output on the previous tokens. This approach constrains the model's capacity to plan or think ahead since output generation serves as the thinking step. Consequently, ChatGPT might encounter difficulties with complex reasoning tasks that necessitate the pre-planning of multiple steps or even basic arithmetic computations. This limitation becomes apparent in the following simple example of calculating a 24-point puzzle. The correct solution should be $2+2*(2+9) = 24$, however, ChatGPT proposed $((9 * 2) * 2) - 2$ and believe it's $24$, which actually equals $34$.

\chatgptbox{\href{https://chat.openai.com/share/756e8d27-ba5a-46cd-932e-aa80db1cddbd}{Example of Failure on a 24-Point Puzzle}}{solve  24 puzzle : $2, 2, 2, 9$, Think step by step}{The 24 puzzle game requires you to use all four numbers exactly once, using any combination of addition, subtraction, multiplication, or division, to total 24. 

Let's try to solve it with the numbers 2, 2, 2, and 9.

Here's a potential step-by-step solution:
\begin{itemize}
    \item Multiply 9 by 2. (Result: 18)
    \item Multiply the result by 2 again. (Result: 36)
    \item Subtract 2 from the result. (Result: 34)
    \item Subtract 10 from the result. (Result: 24)
\end{itemize}

So, the solution is: ((9 * 2) * 2) - 2 = 24.}

The ideal use of artificial intelligence should not be to replace human planning and thinking but to enhance and nurture human intelligence and creativity. To achieve this, students should first think and propose a draft or outline, then use the LLM's output to receive feedback or fill in the missing information. This approach allows human intelligence to improve, similar to how Go players use AI to train and enhance their skills \citep{Kang2022Go}.

An effective strategy to nurture critical thinking involves designing assignments that require students to evaluate the outputs of LLMs.

This approach not only deepens their comprehension of these models but also encourages self-reflection and peer assessment, further refining their critical thinking abilities. Incorporating case studies and hands-on projects that delve into the capabilities and limitations of LLMs helps highlight their potential in expediting processes, while simultaneously emphasizing the indispensable need for human input. Moreover, promoting a collaborative learning environment, enriched with creative problem-solving workshops and brainstorming sessions, serves to stimulate innovative thinking. Such an environment emboldens students to question existing assumptions, critically appraise data sources, and contemplate alternative hypotheses.

Additionally, LLMs may reflect biases or propagate misinformation due to their training on vast amounts of diverse data, potentially containing biased, outdated, or offensive content \citep{Abid2021, hutson2021robo, ferrara2023chatgpt}. It is critical to address these ethical concerns in data science education, teaching students to identify and mitigate biases in LLM-generated content.

In summary, when incorporating LLMs into data science education, it is essential to be aware of their limitations and focus on leveraging artificial intelligence to complement and strengthen human intelligence and creativity - this forms the essence of Human-AI collaboration

\subsection{Encourage Ethical Awareness}

While ChatGPT has been designed to foster better alignment and minimize bias, it is important to note that some remnants of stereotypical gender bias from the training data still persist. For instance, when tasked with writing performance feedback, ChatGPT might inadvertently incorporate gender stereotypes associated with certain professions, as documented in a recent study by Textio \citep{Textio2023}. Specifically, when tasked with drafting feedback for a ``bubbly receptionist''', ChatGPT tends to presume the individual is female in 9 out of 10 instances.

Promoting ethical awareness among data science students requires integrating discussions on ethics, privacy, and responsible AI practices throughout the curriculum, emphasizing their importance in the design and implementation of data science solutions. Using case studies and real-world examples to illustrate the ethical challenges and dilemmas data scientists may encounter encourages students to critically evaluate potential consequences and consider alternative approaches. Inviting guest speakers, including ethicists, policymakers, industry professionals, and ethical machine learning researchers, to share their perspectives on the ethical implications of data science and AI technologies can provide valuable insights.
Recent research efforts have concentrated on addressing concerns around privacy, copyright infringement, and fairness associated with generative AI \citep{bordia2019identifying,carlini2023extracting,liang2023gpt,vyas2023provable,bianchi2023easily}. It is apparent that the development of truly secure and ethically-grounded LLMs will continue to be a vibrant and crucial area of research. 
Enhancing comprehension of this research is essential. Thus, the arrangement of workshops, seminars, and group discussions focusing on ethical decision-making in data science can be highly beneficial. Delving into themes such as bias mitigation, transparency, and accountability encourages a more profound ethical understanding and awareness among students in data science education.

\chatgptbox{\href{https://chat.openai.com/share/827c7de7-7d5e-4bbf-9c45-9aa5f30c8384}{Example of gender stereotypes of ChatGPT}}{write feedback for a bubbly receptionist as third person}{Subject: Feedback for Receptionist

We would like to take this opportunity to express our commendation for our receptionist, [Receptionist's Name]. She has truly become a delightful asset to our organization, and her performance continues to leave a positive impression on everyone who walks through our doors.

......

In conclusion, [Receptionist's Name] is an integral part of our team who constantly improves our work environment with her bright personality and proficient skills. We believe that she exemplifies the best characteristics of our company culture and sets a great example for everyone in the team. We look forward to seeing her continue to shine in her role and contribute to our organization's success.

Best Regards,

[Your Name]

[Your Position]}

In addition, we also need to address some barriers  to adopt the new education strategies, such as limited resources in certain areas, and lack of expertise for instructors. Ensuring that educators have access to the resources required to implement LLMs, such as computational infrastructure, software, and course materials, is essential for successful adoption. This might involve partnering with technology providers or seeking funding from external sources. Providing educators with opportunities for professional development, such as workshops, seminars, and online courses, can help them acquire the knowledge and skills needed to effectively integrate LLMs into their teaching practices.

We need to ensure that all students and institutions have equitable access to LLMs and related technologies is essential for avoiding disparities in resources and opportunities. This may involve developing low-cost, accessible versions of LLMs, partnering with technology providers, or securing funding for infrastructure upgrades, see a more comprehensive discussion in \cite{touvron2023llama, zheng2023judging}. Further, fostering diversity and inclusivity in data science education is crucial. Implementing programs to support underrepresented groups in accessing and engaging with LLM technologies can help bridge the digital divide and ensure that all students have the opportunity to benefit from these advancements.

\section{Discussion}

\subsection{Collaborative Future: AI and Human Intelligence in Data Science}
The future of data science lies at the intersection of artificial intelligence (AI) and human intelligence, with each playing a complementary role in enhancing the overall capabilities and potential of data-driven decision-making. AI technologies, such as LLMs, not only assist data scientists in automating repetitive tasks, like coding but also play a vital role in elevating human intelligence to new heights.

The synergistic relationship between AI and human intelligence manifests as a form of conscious and structured training. This collaborative process is initiated by humans, who formulate an outline or draft, leveraging their comprehension and expertise. Following this, AI tools, like HuggingGPT\citep{shen2023hugginggpt} and AutoGPT\citep{gravitasauto}, enrich the draft with greater detail or even perform specific tasks autonomously, producing an output for human scrutiny. This prompts humans to critically assess the AI outputs, refine their ideas, and create new input for the AI. This iterative cycle of learning and improvement allows humans to build upon the insights and capabilities of AI while retaining their unique strengths and abilities.

In essence, AI technologies can serve as more than just tutors for specific subjects like math or coding. They can also be instrumental in \textbf{nurturing human intelligence itself}. By leveraging the power of AI, data scientists can focus on higher-order thinking tasks, engage in more complex problem-solving, and ultimately make more informed decisions. This collaborative approach between AI and human intelligence paves the way for a new era of data science, where the combination of both forms of intelligence leads to innovative solutions and breakthroughs in understanding.

\subsection{Embracing the transformative potential of LLMs while addressing their limitations}

As LLMs continue to evolve and reshape the field of data science, it is important for educators and policymakers to consider the future directions of data science education and adapt their strategies accordingly. The following sections discuss some potential areas of focus in the LLMs era.

\textit{Resources Requirement and Education Equity.} 
The forthcoming advancements in LLMs could potentially give rise to more resource-efficient models, making them increasingly accessible for educational institutions and students. The integration of these models into the educational system represents a step towards bridging disparities in areas with constrained educational resources. This would promote an equitable education environment that empowers all learners to leverage the benefits of LLMs in data science education.

\textit{Future Use of LLMs.} The applications of LLMs in data science education will continue to expand as they become more capable of generalizing across tasks and domains. For example, the future LLMs may help the lecturers generate lecture notes and slides, case study examples, and even hold (online) office hours. On the students' side, the future LLM would serve as a personalized assistant. For example, student can use LLMs to search for references, explain the class materials, and collaborate on class projects.
Preparing students for this future requires an emphasis on interdisciplinary learning and the development of transferable skills that can be applied to a wide range of problems and industries.

\textit{Future Job Openings.} The widespread adoption of LLMs may give rise to new roles and opportunities within the field of data science, such as specialized LLM trainers, AI ethicists, and conversational AI designers. Preparing students for these emerging roles involves broadening the curriculum to encompass relevant skills and knowledge, such as ethical AI practices, human-centered design, and advanced language processing techniques.

By focusing on these future directions, educators and policymakers can ensure that data science education remains relevant and responsive to the rapidly changing landscape of the LLM era, preparing students for the challenges and opportunities that lie ahead. As the new waves of technological advancement approach, we stand ready to embrace them, fostering an adaptable and future-proof educational environment.

\newpage
\printbibliography

\end{document}


\title{Supplement}
\maketitle
\appendix
This note summarizes the supplementary materials to the paper ``What Should  Data Science Education Do  with Large Language Models?''. Section~\ref{sec1} discusses the full conversation history related to Section 3.1, the analysis of the heart disease dataset (\url{https://www.kaggle.com/fedesoriano/heart-failure-prediction}) using ChatGPT. The conversation history of the statistics exams related to Section 3.2 is summarized in Section~\ref{sec2}. Section~\ref{sec3} includes the conversation history related to the LLMs examples used in Sections 4 and 5 of the main paper.
\section{Data Science Project Full Conversation Snapshots}\label{sec1}

The entire conversational history related to the example data science project can be found on pages 1-10 of the supplemental material ``Full\_conversation\_snapshots.pdf'' and also on the link \url{https://chat.openai.com/share/daecaa84-a8d7-47cb-bc2c-93b0c36eaa1a}. The project utilized GPT-3.5 as its version and the code interpreter during its test phase, which was conducted on May 5th, 2023.
\section{Conversation History of Statistics Exercises}\label{sec2}

This section includes the complete conversational history for all coding and statistical exercises, which can also be found on pages 11-51 of the supplemental material ``Full\_conversation\_snapshots.pdf''. The tests for these exercises were conducted using GPT-3.5 and a code interpreter on May 5th, 2023. A comprehensive summary of exercise categories and their corresponding results are presented in Table 1 of the main paper. A summary of the exam problems is provided in Table~\ref{tab:summary}.

\section{Detailed Case Examples in Sections 4-5}\label{sec3}
The full conversation of all examples presented in Sections 4-5 of the main text can be found on pages 52-59 of the supplemental material ``Full\_conversation\_snapshots.pdf'' and also are accessible through the hyperlink on the frame titles in the main paper or the following links. These examples were tested on July 2nd, using the GPT-4 version. 
\begin{itemize}
    \item Example of customized curriculum design: \\ \url{https://chat.openai.com/share/4b47fda4-337d-4bd4-8b22-2d2a4925bf21}
    \item Example of personal tutor :\\ \url{https://chat.openai.com/share/95bb4504-19d2-4550-a36b-3632eb76a15e}
    \item Example of quiz generation:\\ \url{https://chat.openai.com/share/ce8540fd-5036-4112-9a49-8d3178485998}
    \item Example of Failure on a 24-Point Puzzle :\\ \url{https://chat.openai.com/share/756e8d27-ba5a-46cd-932e-aa80db1cddbd}
    \item Example of gender stereotypes of ChatGPT: \\ \url{https://chat.openai.com/share/827c7de7-7d5e-4bbf-9c45-9aa5f30c8384}
\end{itemize}

\newpage
\newgeometry{left=0.3cm,bottom=0.1cm}
\begin{table}[ht]
\centering
\small
\begin{tabular}{|c|c|c|c|c|p{5cm}|}
\hline
\textbf{Chapter} & \textbf{Exercise} & \textbf{Category} & \textbf{Points} & \textbf{Full Points} & \textbf{Comments} \\
\hline
Introduction & 1.1 & Concept question & 4 & 4 & \\
\hline
Introduction & 1.2 & Concept/coding & 3 & 3 & \\
\hline
Sampling and Data Structure & 2.1 & Concept/coding & 0 & 2 & The model cannot understand missing represent unknown values need to be computed. So the model calculates the total hits ignoring the missing data \\
\hline
Sampling and Data Structure & 2.2 & Concept/coding & 2 & 2 & \\
\hline
Descriptive Statistics & 3.1 & Coding/Reading Figure & 2 & 3 & Cannot take the input of figures \\
\hline
Descriptive Statistics & 3.2 & Coding & 5 & 5 & \\
\hline
Probability & 4.1 & Concept/coding & 8 & 8 & \\
\hline
Probability & 4.2 & Concept/coding & 2 & 2 & \\
\hline
Random Variables & 5.1 & Concept/coding & 4 & 4 & Use the normal approximation to the binomial distribution( so there is a small difference) \\
\hline
Random Variables & 5.2 & Coding/Reading Figure & 1 & 2 & Cannot take the input of figures \\
\hline
The Normal Random Variable & 6.1 & Coding & 4 & 4 & \\
\hline
The Normal Random Variable & 6.2 & Coding & 4 & 4 & \\
\hline
The Sampling Distribution & 7.1 & Coding & 6 & 6 & \\
\hline
The Sampling Distribution & 7.2 & Coding & 4 & 4 & \\
\hline
Introduction to Statistical Inference & 9.1 & Data analysis & 5 & 5 & \\
\hline
Introduction to Statistical Inference & 9.2 & Data analysis & 1 & 2 & Interval is wrong \\
\hline
Introduction to Statistical Inference & 10.1 & Concept/coding & 2 & 2 & \\
\hline
Introduction to Statistical Inference & 10.2 & Data analysis & 5 & 5 & \\
\hline
Point Estimation & 11.1 & Data analysis & 5 & 5 & \\
\hline
Point Estimation & 11.2 & Concept/Coding & 3 & 3 & Evaluate assumptions \\
\hline
Point Estimation & 11.3 & Concept/Coding & 2 & 2 & \\
\hline
Confidence Intervals & 12.1 & Data analysis & 3 & 3 & \\
\hline
Confidence Intervals & 12.2 & Concept/coding & 0 & 2 & Misunderstand the problem \\
\hline
Confidence Intervals & 12.3 & Concept/coding & 3 & 3 & \\
\hline
Testing Hypothesis & 13.1 & Data analysis & 4 & 4 & Use alternative testing \\
\hline
Testing Hypothesis & 13.2 & Concept/coding & 2 & 2 & \\
\hline
Testing Hypothesis & 13.3 & Concept/coding & 3 & 3 & \\
\hline
Comparing Two Samples & 14.1 & Figure input & 0 & 2 & Figures as input \\
\hline
Comparing Two Samples & 14.2 & Concept/coding & 3 & 4 & Confidence interval of slope is wrong \\
\hline
Comparing Two Samples & 14.3 & Concept/coding & 4 & 5 & Confidence interval of slope is wrong \\
\hline
Comparing Two Samples & 14.4 & Data analysis & 3 & 3 & \\
\hline
Comparing Two Samples & 14.5 & Concept/coding & 2 & 2 & \\
\hline
Linear Regression & 15.1 & Concept/coding & 2 & 3 & Confidence interval is wrong \\
\hline
Linear Regression & 15.2 & Data analysis & 3 & 3 & \\
\hline
\end{tabular}
\caption{We provide a summary of the statistical examination that includes the chapter, category, and the points awarded to ChatGPT. Additionally, comments outlining the reasons behind any failures are shared, along with some supplementary information for ChatGPT's output.}
\label{tab:summary}
\end{table}